\documentclass[12pt]{elsarticle}                                          
\usepackage{axodraw4j}
\usepackage{hyperref}
\usepackage{graphicx}                                                         
\usepackage{a41}                                                                
\usepackage{xcolor}                     
\usepackage[rflt]{floatflt}
\usepackage{float}
\usepackage{lscape}
\usepackage{hyperref}
\hypersetup{colorlinks=true, 
linkcolor=blue, filecolor=blue, urlcolor=mygreen}
\usepackage{breakurl} 
\usepackage{times}     %
\usepackage{array}
\usepackage[english]{babel}
\usepackage[latin1]{inputenc}
\usepackage[T1]{fontenc}
\usepackage{ae}
\usepackage{url}
\usepackage{amsmath, amsthm, amssymb}
\usepackage{slashed}

\usepackage{rotating}
\usepackage{graphicx}
\usepackage{comment}
\newcounter{mmacnt}
\def\restartmma{\setcounter{mmacnt}{0}}
\restartmma \catcode`|=\active
\def|#1|{\mathrm{#1}}
\catcode`|=12
\newenvironment{mma}{
\par\smallskip
\catcode`|=\active
\parskip=0pt\parindent=0pt 
\small
\def\In##1\\{%
\def\linebreak{\hfill\break\null\qquad}%
\refstepcounter{mmacnt}
\hangindent=2.5em\hangafter=0
\leavevmode
\llap{\tiny\sffamily In[\arabic{mmacnt}]:=\kern.5em}%
\mathversion{bold}\footnotesize$
\displaystyle##1$\normalsize
\mathversion{normal}\par
 }%
\def\Print##1\\{%
\def\linebreak{\hfill\break}%
\hangindent=2.5em\hangafter=0
\leavevmode ##1\par}%
\def\Out##1\\{%
\def\linebreak{$\hfill\break\null\hfill$}%
\kern\abovedisplayskip\par
\hangindent=2.5em\hangafter=0
\leavevmode
\llap{\tiny\sffamily Out[\arabic{mmacnt}]=\kern.5em}
\footnotesize$\displaystyle##1$
\normalsize\hfill\null\par
\kern\belowdisplayskip
}%
\def\Warning##1##2\\{%
\def\linebreak{\hfill\break}%
\hangindent=2.5em\hangafter=0
\leavevmode
{\scriptsize##1 : ##2}\par}%
}{%
\par\smallskip
}

\usepackage{color}
\newenvironment{fshaded}{%
\MakeFramed {\FrameRestore}
}%
{\endMakeFramed}

\makeatletter
\def\ps@pprintTitle{%
\let\@oddhead\@empty
\let\@evenhead\@empty
\def\@oddfoot{\reset@font\hfil\thepage\hfil}
\let\@evenfoot\@oddfoot
}
\makeatother
\usepackage{tikz}
\usetikzlibrary{matrix}
\allowdisplaybreaks[4]

\newcommand{\n}{\nonumber}
\begin{document}
\begin{frontmatter}
\title{\huge 
\textbf{One-loop $W$ boson contributions to the decay 
$H\rightarrow Z\gamma$ in the general $R_\xi$ gauge}}
\author{Dzung Tri Tran}
\address{\it 
University of Science Ho Chi Minh City, $227$ 
Nguyen Van Cu, District $5$, HCM City, Vietnam}
\author{Le Tho Hue}
\address{\it Institute of Physics, Vietnam Academy of Science and Technology, 
$10$ Dao Tan, Ba Dinh, Hanoi, Vietnam}
\author{Khiem Hong Phan}
\ead{phanhongkhiem@duytan.edu.vn}
\address{\it Institute of Fundamental and Applied Sciences, 
Duy Tan University, Ho Chi Minh City $700000$, Vietnam\\ 
Faculty of Natural Sciences, Duy Tan University, 
Da Nang City $550000$, Vietnam}
\pagestyle{myheadings}
\markright{}
\begin{abstract} 
 One-loop $W$ boson contributions to 
the decay $H\rightarrow Z\gamma$ in 
the general $R_\xi$ gauge are presented. 
The analytical results are expressed
in terms of well-known Passarino-Veltman functions
which their numerical evaluations
can be generated using {\tt LoopTools}. 
In the limit $d\rightarrow 4$,  we have shown that 
these  analytical results  are independent of the unphysical
parameter $\xi$ and consistent with previous results. 
The gauge parameter independence are also checked 
numerically for consistence. Our results are also 
well stable with different values of 
$\xi =0, 1, 100,$ and $\xi \rightarrow \infty$. 
\end{abstract}
\end{frontmatter}
\section{Introduction}
\noindent
The decay process of the standard model-like 
(SM-like) Higgs boson $H\rightarrow Z\gamma$ is of great  interest 
at the Large Hadron Collider (LHC) as well as future 
colliders~\cite{Abazov:2008wg,Chatrchyan:2013vaa,Aaboud:2017uhw,Aad:2020plj}. 
Similar to the important loop-induced decay $H\rightarrow \gamma\gamma$, 
which is one of the key channel for finding the SM-like Higgs boson at LHC, 
the partial decay width of the decay $H\rightarrow Z\gamma$ 
will provide important information on the nature of Higgs sector.
Since the leading contributions to this decay amplitude are from 
one-loop Feynman diagrams, it is sensitive to new physics predicted 
by many recent models beyond the standard model (BSM), i.e., 
new contributions of many new heavy charged particles that exchange 
in the loop diagrams. Therefore, detailed calculations for one-loop 
and higher-loop contributions to the decay channel 
$H\rightarrow Z\gamma$ are necessary.

There have been many computations for one-loop 
contributions to 
the decay channel $H\rightarrow Z\gamma$
within standard model (SM) and its extensions 
in~\cite{Cahn:1978nz,Bergstrom:1985hp,Martinez:1989kr,Djouadi:1996yq,Djouadi:1997yw,Chiang:2012qz,Chen:2013vi,Cao:2013ur,Bonciani:2015eua,Hammad:2015eca,Belanger:2014roa,No:2016ezr,Monfared:2016vwr,Fontes:2014xva,Funatsu:2015xba,Hue:2017cph,Hung:2019jue,Herrero:2020dtv,Dedes:2019bew,Gehrmann:2015dua}
also in the references therein. In the paper Ref.~\cite{Boradjiev:2017khm}, 
the authors have proposed the dispersion theoretic evaluations for $H\rightarrow Z\gamma$. 
In addition, hypergeometric  presentation for one-loop 
contribution to the amplitude of the decay  $H\rightarrow Z\gamma$
has been presented in Ref. \cite{Phan:2020xih}. 
Almost the calculations were carried out in the unitary gauge 
because of the less number of the  Feynman diagrams
in this gauge than the other ones. However, 
the results may appear problems of the 
large numerical cancellations, especially  the higher-rank tensor 
one-loop integrals occur in the diagrams due to the 
$W$ boson exchange.  In our opinion,  the derivation the  one-loop 
$W$ boson contributions  to the decay amplitude  $H\rightarrow Z\gamma$
in the general $R_{\xi}$ gauge is mandatory, even in the SM framework. This helps
to verify the correctness of the final results 
 supposed
to be independent of  the unphysical  parameter $\xi$. 
Furthermore, one can obtain a good stability
of the results by fixing suitable values of $\xi$. 

Many recent BSMs are electroweak gauge extensions such as the left-right models (LR) constructed from the $SU(2)_L\times SU(2)_R\times U(1)_Y$~\cite{Pati:1974yy, Mohapatra:1974gc, Senjanovic:1975rk},  the 3-3-1 models ($SU(3)_L\times U(1)_X$)~\cite{Singer:1980sw, Valle:1983dk, Pisano:1991ee, Frampton:1992wt, Diaz:2004fs,Foot:1994ym}, the  3-4-1 models ($SU(4)_L\times U(1)_X$)~\cite{Foot:1994ym}, ect.  They all predict new charged gauge bosons which may give considerable one-loop contributions    to the decay amplitude  $H\rightarrow Z\gamma$.  Once their couplings and the respective  golstone bosons and ghosts    are determined,  their contributions to the decay amplitude $H\rightarrow Z\gamma$ can be presented analytically using the results given in this paper, although it is limited in the standard model framework. They can be used to cross-check with other results calculated in the unitary gauge~\cite{Hue:2017cph}.  This is another way to confirm the complicated  properties of the couplings relating with new goldstone bosons appearing in BSM. 

As the above reasons, detailed calculations for one-loop $W$ 
boson contributions to 
$H\rightarrow Z\gamma$ in the $R_\xi$ gauge  will be presented in this paper. 
The analytical results will be grouped in  form factors that are  written 
in terms of the Passarino-Veltman functions
so that their numerical evaluations
can be generated by  {\tt LoopTools}~\cite{Hahn:1998yk}.   In the limit  
$d\rightarrow4$,  the  analytic results  will be used to  check for 
the $\xi$-independence
and confirm previous results.  Numerical checks
for the $\xi$-independence of the form factors
will be  also discussed. The stability of  results  will be tested  
with varying $\xi =0, 1, 100$ and 
$\xi \rightarrow \infty$. 

The layout of the paper is as follows: In section 2,
we present briefly one-loop tensor reduction method. 
Notations  for one-loop form factors contributing to the amplitude 
of the  SM-like Higgs decay into a $Z$ boson and a photon will 
be defined before  listing all analytical results  in this section. 
Conclusions and outlook are devoted in section $3$. In appendices, 
Feynman rules and one-loop amplitude for the decay channel are 
discussed. 
\section{ \label{sec_Cal}Calculations}   
\noindent
In general, an one-loop decay amplitude is  decomposed 
into one-loop tensor integrals which can be reduced 
frequently to the final forms being sums of only scalar functions. 
Our calculation will follow the tensor reduction method 
for one-loop integrals developed in Ref.~\cite{Denner:2005nn}. 
This technique is described briefly in the following.  

The notations of one-loop one-, 
two- and three-point tensor integrals with rank $P$
are given by
\begin{eqnarray}
 \{A; B; C\}^{\mu_1\mu_2\cdots \mu_P}= \int \frac{d^dk}{(2\pi)^d} 
 \dfrac{k^{\mu_1}k^{\mu_2}\cdots k^{\mu_P}}{\{D_1; D_1 D_2; D_1D_2D_3\}}.
\end{eqnarray}
In this formula, 
$D_j$ ($j=1,2,3$) are the inverse Feynman propagators
\begin{eqnarray}
 D_j = (k+ q_j)^2 -m_j^2 +i\rho,
\end{eqnarray}
$q_j = \sum\limits_{i=1}^j p_i$, 
$p_i$ are the external momenta,  and  
$m_j$ are internal masses in the loops.
 
The explicit reduction formulas for 
one-loop one-, two-, three-points tensor integrals 
up to rank $P=3$ are written as 
follows~\cite{Denner:2005nn}:
\begin{eqnarray}
A^{\mu}        &=& 0, \\
A^{\mu\nu}     &=& g^{\mu\nu} A_{00}, \\
A^{\mu\nu\rho} &=& 0,\\
B^{\mu}        &=& q^{\mu} B_1,\\
B^{\mu\nu}     &=& g^{\mu\nu} B_{00} + q^{\mu}q^{\nu} B_{11}, \\
B^{\mu\nu\rho} &=& \{g, q\}^{\mu\nu\rho} B_{001} 
+ q^{\mu}q^{\nu}q^{\rho} B_{111}, 
\\
C^{\mu}        &=& q_1^{\mu} C_1 + q_2^{\mu} C_2 
 = \sum\limits_{i=1,2}q_i^{\mu} C_i, \\
 C^{\mu\nu}    &=& g^{\mu\nu} C_{00} 
 + \sum\limits_{i,j=1,2}q_i^{\mu}q_j^{\nu} C_{ij},
\\
C^{\mu\nu\rho} &=&
	\sum_{i=1}^2 \{g,q_i\}^{\mu\nu\rho} C_{00i}+
	\sum_{i,j,k=1}^2 q^{\mu}_i q^{\nu}_j q^{\rho}_k C_{ijk}.
\end{eqnarray}
For convenience, another short notation~\cite{Denner:2005nn} 
$\{g, q_i\}^{\mu\nu\rho}$ will be used 
as follows: $\{g, q_i\}^{\mu\nu\rho} = g^{\mu\nu} q^{\rho}_i 
+ g^{\nu\rho} q^{\mu}_i + g^{\mu\rho} q^{\nu}_i$. 
Following this approach, the scalar coefficients 
$A_{00}, B_1, \cdots, C_{222}$
in the right hand sides of the above equations 
are so-called Passarino-Veltman functions~\cite{Denner:2005nn}. 
Their analytic formulas for numerical calculations are 
well-known. More convenience, these functions can be calculated 
numerically using the available package 
{\tt LoopTools}~\cite{Hahn:1998yk}.

The above notations will be used to  evaluate the  
one-loop $W$  contributions to the  decay amplitude 
$H\rightarrow Z(p_1)\gamma(p_2)$. In the SM framework, 
these contributions comes from the Feynman diagrams 
given in Fig.~\ref{Wboson}, where all particles $W$ boson, 
Goldstone boson and Ghost exchanging in the loop 
must be considered in the general $R_{\xi}$ gauge.

The total amplitude 
of the decay channel is then expressed
in terms of the Lorentz invariant structure 
as follows:
\begin{eqnarray} \label{eq_AHZga}
 \mathcal{A}_{H\rightarrow Z \gamma} =
 \mathcal{A}_{\mu\nu} 
 \epsilon_\mu^*(p_1)
 \epsilon_\nu^*(p_2)=
 \Big\{\mathcal{A}_{00} g^{\mu\nu} + \sum\limits_{i,j=1}^2 \mathcal{A}_{ij}
 p_i^{\mu}p_j^{\nu}  
 \Big\} \epsilon_\mu^*(p_1)
 \epsilon_\nu^*(p_2). 
\end{eqnarray}
All kinematic invariant variables 
are relevant in this process:   
\begin{eqnarray}
p_1^2&=&  M_Z^2,\\
p_2^2&=&0,\\
p^2 &=& (p_1+p_2)^2= M_H^2,
\end{eqnarray}
which results in a consequence that 
\begin{eqnarray}
p_1p_2 = \frac{M_H^2-M_Z^2}{2}.
\end{eqnarray}
The Ward identity $p_2^{\nu}\epsilon_\nu^*(p_2) =0$ implies 
that the two form factors $\mathcal{A}_{22}$ 
and $\mathcal{A}_{12}$ do not contribute 
to the amplitude given in Eq.~\eqref{eq_AHZga}. 
In addition, we have $p_2^{\nu}\mathcal{A}_{\mu\nu} =0$, 
leading to another zero form factor, namely $\mathcal{A}_{11}=0$. 
Now, the amplitude has a very simple form as follows  
\begin{eqnarray}
\label{formfactor}
 \mathcal{A}_{H\rightarrow Z \gamma} =
 \Big\{\mathcal{A}_{00} g^{\mu\nu} + \mathcal{A}_{21}
 p_1^{\nu}p_2^{\mu}  
 \Big\} \epsilon_\mu^*(p_1)
 \epsilon_\nu^*(p_2). 
\end{eqnarray}
The form factors $\mathcal{A}_{00}, \mathcal{A}_{21}$
will be  expressed in terms of the Passarino-Veltman functions 
mentioned in the beginning of this section. 
The derivation  
are performed with the help of {\tt Package-X}~\cite{Patel:2015tea}
for handling all Dirac traces in $d$ dimensions. 
One-loop form factors are presented in the 
standard notations defined in {\tt LoopTools}~\cite{Hahn:1998yk} 
on a diagram-by-diagram basis.
\subsection{In the general $R_{\xi}$ gauge}   
We first arrive the calculations in general $R_{\xi}$ gauge. 
To simplify the computations, the $W$ boson propagator is 
decomposed into the following form
\begin{eqnarray}
\label{WProb}
 \dfrac{- i}{p^2 - M_W^2} 
\Bigg[ g^{\mu \nu} - (1 - \xi) 
\dfrac{p^\mu p^\nu}{p^2 - M_{\xi}^2}  \Bigg]
= \dfrac{- i}{p^2 - M_W^2} \left( g^{\mu\nu} - \dfrac{k^{\mu}k^{\nu}}{M_W^2} \right)
 + \dfrac{-i}{p^2 - M_{\xi}^2}\dfrac{k^{\mu}k^{\nu}}{M_W^2}
\end{eqnarray}
with $M_{\xi}^2= \xi M_W^2$. The first term in the 
right hand side of  Eq.~\eqref{WProb} is nothing
but the $W$ boson propagator in unitary gauge. While the second
term  relates to the propagators of Goldstone boson and Ghost particles. 
In the convention of Eq.~(\ref{WProb}), each diagram with $W$ boson 
exchanging in the loop will be separated into several parts. 
For an example, the Feynman amplitude 
for diagram $(a)$ in Fig.~\ref{Wboson} is divided into $8$ terms 
as follows:
\begin{eqnarray}
 \mathcal{A}^{(a)} = \sum\limits_{i,j,k=1}^{2} \mathcal{A}^{(a)}_{ijk}.
\end{eqnarray}
The notation $\mathcal{A}^{(a)}_{ijk}$ is corresponding to 
which term on the right hand side of Eq.~(\ref{WProb}) is
taken. In this scheme, the amplitude in (\ref{formfactor}) 
is presented by mean of 
\begin{eqnarray}
\label{mainform}
 \mathcal{A}_{H\rightarrow Z \gamma} =
 \Big\{ \Big[\sum \limits_{\text{diag} \equiv \{a, \cdots ,j\}}
 \mathcal{A}_{L}^{(\text{diag})}(\xi)\Big]g^{\mu\nu} 
 + 
 \Big[\sum \limits_{\text{diag} \equiv \{a, \cdots ,j\}}
 \mathcal{A}_{T}^{(\text{diag})}(\xi)\Big]
 p_2^{\mu}p_1^{\nu}  
 \Big\} \epsilon_\mu^*(p_1)
 \epsilon_\nu^*(p_2). 
\end{eqnarray}
The terms $\mathcal{A}_{L}^{(\text{diag})}(\xi)$
and $\mathcal{A}_{T}^{(\text{diag})}(\xi)$ 
will be  collected on a diagram-by-diagram basis 
in the following subsections. In this 
article, we show analytic results for 
$\mathcal{A}_{T}^{(\text{diag})}(\xi)$
as examples. 
\subsubsection{Diagrams $a$ and $a'$}%
We first calculate the topologies $(a+a')$ 
having only $W$ boson in the loop 
diagrams (see the two diagrams $a$ and $a'$ in 
Fig.~\ref{Wboson}). The respective form factors denoted  
in Eq.~(\ref{mainform}) are splitted 
into the $8$  pieces, namely
\begin{eqnarray}
\mathcal{A}_{T}^{(a+a')}(\xi)
&=&
\dfrac{g_{HWW} g_{ZWW}
g_{AWW}}{32 \pi^2 M_W^4}
\Big\{
\mathcal{A}_{111}^{T}(\xi)
+ \mathcal{A}_{112}^{T}(\xi)
+ \mathcal{A}_{121}^{T}(\xi)
+ \mathcal{A}_{211}^{T}(\xi)
+ \mathcal{A}_{122}^{T}(\xi)
\nonumber\\
&&
+ \mathcal{A}_{212}^{T}(\xi)
+ \mathcal{A}_{221}^{T}(\xi)
+ \mathcal{A}_{222}^{T}(\xi)
\Big\}.
\end{eqnarray}
All terms in the above equations are presented
in terms of the Passarino-Veltman functions 
as follows:
\begin{eqnarray}
\mathcal{A}_{111}^{T}(\xi)
&=&
(2 M_H^2+4 M_W^2) 
\Big[
B_{11}
+B_1
\Big] (M_H^2,M_W^2,M_W^2)
+4 M_W^2 B_0(M_H^2,M_W^2,M_W^2)
\nonumber\\
&&
+8 M_W^2(4 M_W^2- M_Z^2) 
C_0(M_H^2,M_Z^2,0,M_W^2,M_W^2,M_W^2)
\nonumber \\
&&
+
2\Big[ 2 M_H^2 (2 M_W^2-M_Z^2)
+8 (d-1) M_W^4-4 M_W^2 M_Z^2
\Big]
\times
\nonumber \\
&&\hspace{4.3cm}
\times
[C_{22}
+C_{12}
+C_2](M_Z^2,0,M_H^2,M_W^2,M_W^2,M_W^2),
\\
&&\n \\
\mathcal{A}_{112}^{T}(\xi)
&=&
-4 M_W^2 B_0(M_H^2,M_W^2,M_{\xi}^2)
+(6 M_W^2 M_Z^2-8 M_W^4) 
C_0(M_Z^2,0,M_H^2,M_W^2,M_W^2,M_{\xi}^2)
\nonumber\\
&&
+2 M_W^2 M_Z^2 
C_1(M_Z^2,0,M_H^2,M_W^2,M_W^2,M_{\xi}^2)
+[M_H^2-M_W^2 (\xi-1)]
\times
\nonumber \\
&&
\times  
\Big\{
(3 M_Z^2-4 M_W^2) 
\Big[
C_{22}
+ C_{12}
\Big] (M_Z^2,0,M_H^2,M_W^2,M_W^2,M_{\xi}^2)
\nonumber \\
&& \hspace{1.0cm}
-2 B_{11}(M_H^2,M_W^2,M_{\xi}^2)
\Big\}
+[M_H^2-M_W^2 (\xi-3)] 
\times
\nonumber \\
&&
\times 
\Big[
-2 B_1(M_H^2,M_W^2,M_{\xi}^2)
-(4 M_W^2 - 3 M_Z^2) 
C_2(M_Z^2,0,M_H^2,M_W^2,M_W^2,M_{\xi}^2) \Big],
\\
&&\n \\
\mathcal{A}_{121}^{T}(\xi)
&=&
2 M_H^2 (M_Z^2-M_W^2)
\Big[
C_{22}+C_{12}+C_2
\Big]
(M_Z^2,0,M_H^2,M_W^2,M_{\xi}^2,M_W^2)
\n \\
&&
+4 M_W^2(M_Z^2-M_W^2)
\Big[
C_{22}+C_{12}+C_2+C_1+C_0
\Big]
(M_Z^2,0,M_H^2,M_W^2,M_{\xi}^2,M_W^2)
, \\
&&\n \\
\mathcal{A}_{211}^{T}(\xi)
&=&
[M_W^2 (\xi+1)-M_H^2] \times
\nonumber\\
&& 
\times
\Big\{
2 B_1(M_H^2,M_{\xi}^2,M_W^2)
+(4 M_W^2- 3 M_Z^2) C_1(M_H^2,M_Z^2,0,M_{\xi}^2,M_W^2,M_W^2)
\Big\}
\n \\
&&
+4 M_W^2 (2 M_W^2 - M_Z^2)
C_1(M_Z^2,0,M_H^2,M_{\xi}^2,M_W^2,M_W^2)
\n \\
&&
+ [M_H^2-M_W^2 (\xi-1)] 
\times
\\
&& 
\times
\Big\{
(3 M_Z^2-4 M_W^2)
\Big[
C_{22}+ C_{12}
\Big]
(M_Z^2,0,M_H^2,M_{\xi}^2,M_W^2,M_W^2)
- 2 B_{11}(M_H^2,M_{\xi}^2,M_W^2)
\Big\}
,\n\\
&& \n \\
\mathcal{A}_{122}^{T}(\xi)
&=&
M_Z^2[M_W^2 (\xi-1) - M_H^2] 
\Big[
C_{12}
+C_{11}
\Big] (M_H^2,M_Z^2,0,M_W^2,M_{\xi}^2,M_{\xi}^2)
\n \\
&&
-2 M_Z^2 M_W^2 
\Big[
C_2+C_0
\Big] (M_H^2,M_Z^2,0,M_W^2,M_{\xi}^2,M_{\xi}^2)
\n \\
&&
- M_Z^2 [M_H^2-M_W^2 (\xi-3)] 
C_1(M_H^2,M_Z^2,0,M_W^2,M_{\xi}^2,M_{\xi}^2),
\\
&&\n \\
\mathcal{A}_{212}^{T}(\xi)
&=&
2 (M_H^2-2 M_{\xi}^2) 
\Big[B_{11}
+ B_1
\Big]
(M_H^2,M_{\xi}^2,M_{\xi}^2)
\n \\
&&
+2 (M_W^2-M_Z^2) (M_H^2-2 M_{\xi}^2) 
\Big[
C_{22}+C_{12}+C_2
\Big]
(M_Z^2,0,M_H^2,M_{\xi}^2,M_W^2,M_{\xi}^2)
,\\
&&\n \\
\mathcal{A}_{221}^{T}(\xi)
&=&
M_Z^2 [M_W^2 (\xi+1)-M_H^2] 
C_2(M_Z^2,0,M_H^2,M_{\xi}^2,M_{\xi}^2,M_W^2)
\n \\
&&
+
M_Z^2 [M_W^2 (\xi-1) - M_H^2]
\Big[
C_{22}+ C_{12}
\Big]
(M_Z^2,0,M_H^2,M_{\xi}^2,M_{\xi}^2,M_W^2)
,
\\
&&\n \\
\mathcal{A}_{222}^{T}(\xi) &=& 0. 
\end{eqnarray}
\subsubsection{Diagram $b$}%
We next consider the topology
$b$ having two $W$ boson 
internal lines. 
One-loop form factors read:
\begin{eqnarray}
\mathcal{A}_{11}^{T}(\xi)
&=& 
\Big[
M_H^2 B_{111}
+ 2 B_{001}
+ (M_H^2-M_W^2) B_{11}
+B_{00}
-M_W^2 (B_1 + B_0)
\Big](M_H^2,M_W^2,M_W^2) , 
\n \\
&&\\
\mathcal{A}_{12}^{T}(\xi)
&=& \Big[
M_W^2 
(B_0 + B_1)
-B_{00}
-M_H^2 
(B_{11}+B_{111})
-2B_{001}
\Big](M_H^2,M_W^2,M_{\xi}^2),\\
\mathcal{A}_{21}^{T}(\xi)
&=&\Big[
[M_W^2 (1-\xi)-M_H^2] 
B_{11}
- M_{\xi}^2 
B_1
-M_H^2 
B_{111}
-B_{00}
-2 B_{001}
\Big] (M_H^2,M_{\xi}^2,M_W^2),\\
\mathcal{A}_{22}^{T}(\xi)
&=&\Big[
(M_H^2+ M_{\xi}^2) B_{11}
+ M_{\xi}^2 B_1
+M_H^2 B_{111}
+B_{00} 
+ 2 B_{001}
\Big](M_H^2,M_{\xi}^2,M_{\xi}^2).
\end{eqnarray}
\subsubsection{Diagrams $c$ and $c'$}%
The form factors due to the triangle 
diagrams having two $W$ bosons and a 
Goldstone boson in the loop are next 
considered. They are expressed in the 
same scheme
\begin{eqnarray}
\mathcal{A}_{T}^{(c+c')}(\xi)
&=&
\mathcal{A}_{110}^{T}(\xi)
+ \mathcal{A}_{120}^{T}(\xi)
+ \mathcal{A}_{210}^{T}(\xi)
+ \mathcal{A}_{220}^{T}(\xi).
\end{eqnarray}
The related terms in the above 
equations are shown
\begin{eqnarray}
\mathcal{A}_{110}^{T}(\xi) 
&=&
\dfrac{g_{HWX} 
g_{ZWW} g_{AWX}}{8 \pi^2 M_W^4}
\Big\{
2 M_W^2 (2 M_W^2-M_Z^2) 
C_0(M_Z^2,0,M_H^2,M_W^2,M_W^2,M_{\xi}^2)
\n \\
&&
-2 M_W^2 M_Z^2 C_1(M_Z^2,0,M_H^2,M_W^2,M_W^2,M_{\xi}^2)
\n \\
&&
+(2 M_W^2-M_Z^2) [M_H^2-M_W^2 (\xi-1)] 
\Big[
C_{12}
+ C_{22}
\Big] (M_Z^2,0,M_H^2,M_W^2,M_W^2,M_{\xi}^2)
\n\\
&&
+(2 M_W^2-M_Z^2) [M_H^2-M_W^2 (\xi-3)] 
C_2(M_Z^2,0,M_H^2,M_W^2,M_W^2,M_{\xi}^2)
\Big\}
\n \\
&& 
+ \dfrac{g_{HWX}
g_{AWW} g_{ZWX}}{8 \pi^2 M_W^4}
\Big\{
4 M_W^4 C_0(M_H^2,M_Z^2,0,M_W^2,M_{\xi}^2,M_W^2)
\n \\
&&
+2 M_W^2 [M_H^2-M_W^2 (\xi-1)]
\Big[
C_{12}
+ C_{11}\Big](M_H^2,M_Z^2,0,M_W^2,M_{\xi}^2,M_W^2)
\n \\
&&
+2 M_W^2 [M_H^2-M_W^2 (\xi-3)] 
C_1(M_H^2,M_Z^2,0,M_W^2,M_{\xi}^2,M_W^2)
\Big\},\\
&&\n \\
\mathcal{A}_{120}^{T}(\xi) 
&=&
\dfrac{g_{HWX} 
g_{ZWW} g_{AWX}}{8 \pi^2 M_W^4}
\Big\{
(M_W^2-M_Z^2) [M_W^2 (\xi-3)-M_H^2] 
C_2(M_Z^2,0,M_H^2,M_W^2,M_{\xi}^2,M_{\xi}^2)
\n \\
&&
+(M_Z^2-M_W^2) [M_H^2-M_W^2 (\xi-1)] 
\Big[
C_{22}
+
C_{12}
\Big] (M_Z^2,0,M_H^2,M_W^2,M_{\xi}^2,M_{\xi}^2)
\n \\
&&
+2 M_W^2 (M_Z^2-M_W^2) 
\Big[
C_1
+ C_0\Big] (M_Z^2,0,M_H^2,M_W^2,M_{\xi}^2,M_{\xi}^2)
\Big\}
\n \\
&&
+ \dfrac{g_{HWX} g_{AWW} 
g_{ZWX}}{8 \pi^2 M_W^4}
\Big\{
M_W^2 [M_W^2 (\xi-3)-M_H^2] 
C_1(M_H^2,M_Z^2,0,M_W^2,M_{\xi}^2,M_{\xi}^2)
\n \\
&&
-M_W^2 [M_H^2-M_W^2 (\xi-1)] 
\Big[
C_{12}
+ C_{11}\Big]
(M_H^2,M_Z^2,0,M_W^2,M_{\xi}^2,M_{\xi}^2)
\n \\
&&
-2 M_W^4 
\Big[
C_2
+ C_0\Big]
(M_H^2,M_Z^2,0,M_W^2,M_{\xi}^2,M_{\xi}^2)
\Big\},
\\
&&\n \\
\mathcal{A}_{210}^{T}(\xi) 
&=&
\dfrac{g_{HWX} g_{ZWW} 
g_{AWX}}{8 \pi^2 M_W^4}
\Big\{
(M_{\xi}^2 - M_H^2)
(M_W^2 - M_Z^2)
\times
\n \\
&&\hspace{5.0cm} \times
\Big[C_{22}
+ C_{12}
+ C_2
\Big](M_Z^2,0,M_H^2,M_{\xi}^2,M_W^2,M_{\xi}^2)
\Big\}
\n \\
&& 
+ 
\dfrac{g_{HWX} g_{AWW} 
g_{ZWX}}{8 \pi^2 M_W^4}
\Big\{
(M_{\xi}^2 - M_H^2) M_W^2 
\times 
\n \\
&&\hspace{5.0cm} \times
\Big[
C_{12}
+ C_{11}
+ C_1
\Big] (M_H^2,M_Z^2,0,M_{\xi}^2,M_{\xi}^2,M_W^2) 
\Big\},\\
&&\n \\
\mathcal{A}_{220}^{T}(\xi) 
&=&
\dfrac{g_{HWX} g_{ZWW} 
g_{AWX}}{8 \pi^2 M_W^4}
\Big\{
M_Z^2(M_{\xi}^2 - M_H^2) 
\Big[
C_{22}
+C_{12}
+C_2
\Big] (M_Z^2,0,M_H^2,M_{\xi}^2,M_{\xi}^2,M_{\xi}^2) 
\Big\}.
\n
\\
\end{eqnarray}
\subsubsection{Diagrams $d$ and $d'$}%
We are now going to consider one-loop two-point 
diagrams with exchanging a $W$ boson and a 
Goldstone boson in the loop (seen diagrams $d$ 
and $d'$). The form factors are 
divided into two parts as follows:
\begin{eqnarray}
\mathcal{A}_{T}^{(d+d')}(\xi) &=&
\mathcal{A}_{10}^{T}(\xi) + \mathcal{A}_{20}^{T}(\xi).
\end{eqnarray}
All components in the equations are given
\begin{eqnarray}
\mathcal{A}_{10}^{T}(\xi) &=&0,
\\
\mathcal{A}_{20}^{T}(\xi) 
&=&
\dfrac{g_{HZWX} 
g_{AWX}}{8 \pi^2 M_W^2}
\Big[ B_{00}
-M_W^2 B_0\Big] (0,M_W^2,M_{\xi}^2) 
\n \\
&&
+\dfrac{g_{HAWX} 
g_{ZWX}}{8 \pi^2 M_W^2}
\Big[
B_{00}
-M_W^2 B_0\Big]
(M_Z^2,M_{\xi}^2,M_W^2) .
\end{eqnarray}
\subsubsection{Diagrams $e$ and $e'$}%
One-loop topologies with two Goldstone bosons
and one $W$ boson in internal lines are concerned.
The form factors are written as:
\begin{eqnarray}
\mathcal{A}^{(e+e')}_T(\xi) &=&
\dfrac{g_{HWX} 
(g_{ZWX} g_{AXX}+g_{AWX} 
g_{ZXX})}{4 \pi^2 M_W^2}
\Big[
\mathcal{A}_{100}^{T}(\xi)+ 
\mathcal{A}_{200}^{T}(\xi)
\Big].
\end{eqnarray}
The related terms in the equation are decomposed as
\begin{eqnarray}
\mathcal{A}_{100}^{T}(\xi) 
&=&
[ M_H^2 + M_W^2 (3 - \xi) ] C_2(M_Z^2,0,M_H^2,M_W^2,M_{\xi}^2,M_{\xi}^2)
\\
&&
+ 2 M_W^2 \Big[
C_0
+ C_1
\Big] (M_Z^2,0,M_H^2,M_W^2,M_{\xi}^2,M_{\xi}^2)
\n \\
&&
+ [ M_H^2 + M_W^2 (1 - \xi) ]
\Big[
C_{22}
+ C_{12}\Big]
(M_Z^2,0,M_H^2,M_W^2,M_{\xi}^2,M_{\xi}^2),
\n \\
&&\n \\
\mathcal{A}_{200}^{T}(\xi) 
&=&
(M_{\xi}^2 -M_H^2)
\Big[
C_{22}
+ C_{12}
+ C_2
\Big] (M_Z^2,0,M_H^2,M_{\xi}^2,M_{\xi}^2,M_{\xi}^2).
\end{eqnarray}
\subsubsection{Diagrams $f$ and $f'$}
Other topologies with two $W$ bosons
and one Goldstone in the loop are 
mentioned. The corresponding form 
factors are presented in the form of
\begin{eqnarray}
\mathcal{A}_{T}^{(f+f')}(\xi) 
&=&
\dfrac{g_{HWW} 
g_{ZWX} g_{AWX}}{16 \pi^2 M_W^4}
\Big[
\mathcal{A}_{101}^{T}(\xi)
+ \mathcal{A}_{102}^{T}(\xi)
+ \mathcal{A}_{201}^{T}(\xi)
+ \mathcal{A}_{202}^{T}(\xi)
\Big].
\end{eqnarray}
Where the relevant terms are expressed in terms 
of Passarino-Veltman functions. The results read 
in detail 
\begin{eqnarray}
\mathcal{A}_{101}^{T}(\xi) 
&=&
 (M_H^2+2 M_W^2) 
\Big[
C_{22}
+C_{12}
+C_2
\Big](M_Z^2,0,M_H^2,M_W^2,M_{\xi}^2,M_W^2)
\n \\
&&
+2 M_W^2 
\Big[
C_0
+C_1
\Big](M_Z^2,0,M_H^2,M_W^2,M_{\xi}^2,M_W^2),
\\
&&\n \\
\mathcal{A}_{102}^{T}(\xi) 
&=&
 [M_W^2 (\xi-1)-M_H^2] 
\Big[
C_{22}
+C_{12}
\Big](M_Z^2,0,M_H^2,M_W^2,M_{\xi}^2,M_{\xi}^2)
\n \\
&&
-2 M_W^2 
\Big[
C_0
+C_1
\Big]
(M_Z^2,0,M_H^2,M_W^2,M_{\xi}^2,M_{\xi}^2)
\n \\
&&
+ [M_W^2 (\xi-3) - M_H^2] 
C_2(M_Z^2,0,M_H^2,M_W^2,M_{\xi}^2,M_{\xi}^2)
, \\
&&\n \\
\mathcal{A}_{201}^{T}(\xi) 
&=& 
 [M_W^2 (\xi-1) - M_H^2]
\Big[
C_{22}
+C_{12}
\Big](M_Z^2,0,M_H^2,M_{\xi}^2,M_{\xi}^2,M_W^2)
\n \\
&&
+ [M_W^2 (\xi+1) - M_H^2]
C_2(M_Z^2,0,M_H^2,M_{\xi}^2,M_{\xi}^2,M_W^2)
,\\
&&\n \\
\mathcal{A}_{202}^{T}(\xi) 
&=&
(M_H^2-2 M_{\xi}^2) 
\Big\{
C_{22}
+C_{12}
+C_2
\Big\}(M_Z^2,0,M_H^2,M_{\xi}^2,M_{\xi}^2,M_{\xi}^2).
\end{eqnarray}
\subsubsection{Diagrams $g$ and $g'$}%
Applying the same procedure, the form factors
for diagrams $g$ and $g'$ are shown in this subsection.
The results read
\begin{eqnarray}
\mathcal{A}_{T}^{(g+g')}(\xi) 
&=&
\dfrac{g_{HXX} g_{ZWX} g_{AWX}}{8 \pi^2 M_W^2}
\Big[
\mathcal{A}_{010}^{T}(\xi) + \mathcal{A}_{020}^{T}(\xi) \Big].
\end{eqnarray}
All terms in these equations are obtained
\begin{eqnarray}
\mathcal{A}_{010}^{T}(\xi) 
&=&
\Big[
C_{22}
+C_{12}
+C_2
\Big](M_Z^2,0,M_H^2,M_{\xi}^2,M_W^2,M_{\xi}^2)
,
\\
\mathcal{A}_{020}^{T}(\xi) 
&=&
-\Big[
C_{22}+ C_{12}+C_2
\Big](M_Z^2,0,M_H^2,M_{\xi}^2,M_{\xi}^2,M_{\xi}^2).
\end{eqnarray}
\subsubsection{Diagrams $h$ and $h'$}
We also have
\begin{eqnarray}
\mathcal{A}_T^{(h+h')}(\xi)
&=& 
\dfrac{g_{HXX} g_{ZXX} g_{AXX}}{2 \pi^2}
\Big[
C_{22}
+C_{12}
+C_2
\Big](M_Z^2,0,M_H^2,M_{\xi}^2,M_{\xi}^2,M_{\xi}^2)
.
\end{eqnarray}
\subsubsection{Diagram $i$}
We next have
\begin{eqnarray}
\mathcal{A}_T^{(i)}(\xi) = 0.
\end{eqnarray}
\subsubsection{Diagrams $j$ and $j'$}%
Finally, we obtain 
\begin{eqnarray}
\mathcal{A}_T^{(j+j')}(\xi) 
&=& -\xi\; 
\dfrac{g_{Hcc} g_{Zcc} g_{Acc} }{4 \pi^2}
\Big[
C_{22}
+C_{12}
+C_2
\Big] (M_Z^2,0,M_H^2,M_{\xi}^2,M_{\xi}^2,M_{\xi}^2).
\end{eqnarray}
\subsection{In 't Hooft-Veltman gauge}  
Summing all of the contributions  listed in the previous subsection,
we obtain the analytic results of the one-loop form factors needed 
to determine the decay amplitude $H\rightarrow Z\gamma$ in the  
general $R_{\xi}$. In this subsection, we set $\xi=1$ corresponding 
to the 't Hooft-Veltman gauge. The 
form factors read in a compact form as follows: 
\begin{eqnarray}
(16\pi^2)\times 
\mathcal{A}_{21}^{T} &=&
\Big\{ 
4 g_{ZWW} [(2d - 3) g_{AWW} g_{HWW} 
+g_{AWX} g_{HWX}]
-4 g_{Acc} g_{Hcc} g_{Zcc}
\n \\
&&
+8 g_{AXX} (g_{HWX} g_{ZWX}+g_{HXX} g_{ZXX}) 
\Big\}
[C_{22} + C_{12}] (M_Z^2,0,M_H^2,M_W^2,M_W^2,M_W^2) 
\n \\
&&
+\Big\{4 
g_{ZWW} [(2 d - 3) g_{AWW} g_{HWW} 
+3 g_{AWX} g_{HWX}]
-4 g_{Acc} g_{Hcc} g_{Zcc}
\n \\
&&
+8 g_{AXX} (3 g_{HWX} g_{ZWX}
+ g_{HXX} g_{ZXX}) 
\Big\}
C_2(M_Z^2,0,M_H^2,M_W^2,M_W^2,M_W^2) 
\n \\
&&
+\Big(
2 g_{AWW} g_{HWW} g_{ZWW}
-8 g_{AWW} g_{HWX} g_{ZWX}
+16 g_{AWX} g_{HWX} g_{ZXX}
\Big) 
\times
\n \\
&&
\times 
C_2(M_H^2,M_Z^2,0,M_W^2,M_W^2,M_W^2) 
\n \\
&&
+\Big(
10 g_{AWW} g_{HWW} g_{ZWW}
+8 g_{AWX} g_{HWX} g_{ZWW}
+16 g_{AXX} g_{HWX} g_{ZWX}
\Big)
\times
\n \\
&&
\times
C_0(M_Z^2,0,M_H^2,M_W^2,M_W^2,M_W^2).
\end{eqnarray}
\subsection{In unitary gauge}          
In the unitary gauge, we only take 
$\mathcal{A}^{(a+a')}_{111}$ and 
$\mathcal{A}^{(b)}_{11}$ 
into account. The result reads
\begin{eqnarray}
(32\pi^2 M_W^4)\times 
\mathcal{A}_{21}^T &=&
\Big[
2 (M_H^2 + 2 M_W^2)
g_{AWW} g_{HWW} g_{ZWW} 
+4 (M_H^2 - M_W^2) 
g_{HWW} g_{ZAWW}
\Big]\times
\n\\
&&
\times
B_{11}(M_H^2,M_W^2,M_W^2)
\n \\
&&
+4 M_W^2 g_{HWW} 
\big(g_{AWW} g_{ZWW} 
- g_{ZAWW}\big) 
B_0(M_H^2,M_W^2,M_W^2)
\n \\
&&
+2 g_{HWW}
\Big[(M_H^2 + 2 M_W^2) g_{AWW} g_{ZWW} 
- 2 M_W^2 g_{ZAWW}
\Big]
B_1(M_H^2,M_W^2,M_W^2) 
\n \\
&&
+4 g_{HWW} g_{ZAWW} 
\Big[
M_H^2 B_{111} + B_{00} + 2 B_{001}
\Big] (M_H^2,M_W^2,M_W^2)
\\
&&
+4 g_{AWW} g_{HWW} g_{ZWW} 
\Big[
2 M_W^2 (M_H^2 - M_Z^2)
- M_H^2 M_Z^2
+4 (d-1) M_W^4
\Big] \n\\
&&\times
[C_{22} + C_{12} + C_2 ](M_Z^2,0,M_H^2,M_W^2,M_W^2,M_W^2)
\n 
\\
&&
+8 M_W^2 \big(4 M_W^2 - M_Z^2 \big) 
g_{AWW} g_{HWW} g_{ZWW}
C_0(M_H^2,M_Z^2,0,M_W^2,M_W^2,M_W^2).
\n
\end{eqnarray}
In the limit  $d\rightarrow 4$, the form factors in the 
three different gauges  $R_{\xi}$,
't Hooft-Veltam  and unitary result in  the same simple form given as
follows:
\begin{eqnarray}
\label{dtend4}
\mathcal{A}^{d\rightarrow 4}_{H\rightarrow Z \gamma} &=&
\dfrac{e g^2 \cos \theta_W}{32 \pi ^2 M_H^2 M_W^3 
\big( M_H^2-M_Z^2 \big)^2}
\Big[
2\, p_2^{\mu } p_1^{\nu }
-(M_H^2-M_Z^2) g^{\mu \nu }
\Big]
\times
\n \\
&& \times
\Bigg\{
M_Z^2 
\sqrt{M_H^4-4 M_H^2 M_W^2} 
\Big[M_H^2 \big(2 M_W^2-M_Z^2 \big)+12 M_W^4-2 M_W^2 M_Z^2
\Big] 
\times
\n \\
&& \times
\ln \Bigg[
\dfrac{\sqrt{M_H^4-4 M_H^2 M_W^2}+2 M_W^2-M_H^2}{2 M_W^2}
\Bigg]
\n \\
&& 
+M_H^2 M_W^2 
\Big[
M_H^2 \big(M_Z^2-6 M_W^2 \big)+12 M_W^4+6 M_W^2 M_Z^2-2 M_Z^4
\Big]
\times
\n \\
&&\times
\ln^2 \Bigg[
\dfrac{\sqrt{M_H^4-4 M_H^2 M_W^2}+2 M_W^2-M_H^2}{2 M_W^2}
\Bigg]
\n \\
&&
+M_H^2 \sqrt{M_Z^4-4 M_W^2 M_Z^2} 
\Big[
M_H^2 \big(M_Z^2-2 M_W^2 \big)
+2 M_W^2 \big(M_Z^2-6 M_W^2 \big)
\Big] 
\times
\n \\
&&\times
\ln \Bigg[
\dfrac{\sqrt{M_Z^4-4 M_W^2 M_Z^2}+2 M_W^2-M_Z^2}{2 M_W^2}
\Bigg]
\n \\
&& 
- M_H^2 M_W^2 \Big[
M_H^2 \big(M_Z^2-6 M_W^2 \big)+12 M_W^4+6 M_W^2 M_Z^2-2 M_Z^4
\Big] 
\times
\n \\
&&\times
\ln^2 \Bigg[
\dfrac{\sqrt{M_Z^4-4 M_W^2 M_Z^2}+2 M_W^2-M_Z^2}{2 M_W^2}
\Bigg]
\n \\
&& 
+ 
2 M_H^2 M_W^2 \big( M_H^2 - M_Z^2 \big)^2
+ M_H^2 \big( 12 M_W^4 - M_H^2 M_Z^2 \big) \big( M_H^2 - M_Z^2 \big)
\Bigg\}.
\end{eqnarray}
By taking $M_Z^2\rightarrow 0$ in Eq.~(\ref{dtend4}), we then verify 
again many previous results for $H\rightarrow \gamma\gamma$, take 
\cite{Marciano:2011gm, Phan:2021ywd} as examples. 
For $W$ bosons exchanging in the loop, their masses are 
included the Feynman's prescription as $M_W^2 -i\rho$. 
Therefore all the above logarithmic functions 
are well-defined in complex plane.
\section{Numerical tests for the $\xi$-independence }
Numerical illustrations of   
the form factors relating with the decay amplitude   
$H\rightarrow Z \gamma$  in different gauges are shown 
in  Table~\ref{t_test}. The last line of this Table 
gives the  numerical value of the form factors after taking out 
the coefficient $\frac{eg^2}{16 \pi^2}$. The related masses 
are fixed as follows:  $M_H = 125$ GeV, 
$M_Z = 91.2$ GeV, and $M_W = 80.4$ GeV. 
We find that the results are 
well stable  with different values  $\xi =0, 1, 100$ and 
$\xi \rightarrow \infty$.  
\begin{landscape}
\begin{table}[htp]
\footnotesize
\begin{center}
\begin{tabular}{lllll}  \hline \\
$\textbf{Diagrams} / \xi$ & $\xi \rightarrow 0$ & $\xi = 1$ & 
$\xi = 100$ & $\xi \rightarrow \infty$ \\ \\ \hline \hline
\\
$a$  &  $0.05985185065310632$ & $0.06215891398415416$ & 
$0.3209209899754058$ & $3.045808943133905 \cdot 10^7$ 
\\ & $+ 0.00467528078266156 \,i$ &  &  &  \\
\\
\hline \\
$b$  & $-0.004780425280300761$ & $0$ &  $-0.2507738185381761$ & $-3.045808936065990 \cdot 10^7$ 
\\ & $ - 0.012912138304675886 \,i$ &  &  &  \\ \\
\hline \\
$c$  &  $0.01420681172938953$ & $0.011041675936102476$ &  $0.0011888522641987549$ & $0.0005410399289955083$ 
\\ & $ + 0.00618230306965115 \,i$ &  &  &  \\ \\
\hline \\
$d$  &  $0$ & $0$ &  $0$ & $0$ 
\\ &  &  &  &  \\ 
\hline \\
$e$  &  $0.003381779143108288$ & $0.0004225556213988878$ &  $0.00007343857541803593$ & $0.00006391517156179955$ 
\\ & $ - 0.003941724009915961 \,i$ &  &  &  \\ \\
\hline \\
$f$  &  $-0.005045377867037107$ & $0$ &  $0.0001476405256009515$ & $0.0002619707739754035$ 
\\ & $ + 0.005184705008759242 \,i$ &  &  &  \\ \\
\hline \\
$g$  &  $0.0001233773368155624$ & $0$ &  $1.0436687818979681 \cdot 10^{-6}$ & $1.0553816412933658 \cdot 10^{-14}$ 
\\ & $ + 0.0008115734535198974 \,i$ &  &  &  \\ \\
\hline \\
$h$  &  $0.004721799775923751$ & $-0.002769766130017844$ &  $-0.00001581903805520466$ & $-1.575446161278000 \cdot 10^{-13}$ 
\\ &  &  &  &  \\ 
\hline \\
$i$  &  $0$ & $0$ &  $0$ & $0$ 
\\ &  &  &  & \\ 
\hline \\
$j$  &  $\mathcal{O}(10^{-23})$ & $0.001606436079367905$ &  $0.0009174880578314728$ & $0.0009137426900957666$ 
\\ &  &  &  &  \\ 
\hline \\
Sum  &  $0.07245981549100559$ & $0.07245981549100559$ &  $0.07245981549100559$ & $0.07245981549100559$ 
\\ &  &  &  & \\      
\hline\hline
\end{tabular}
\caption{Numerical checks
for the $\xi$-independence of the
form factors are studied. \label{t_test}}
\end{center}
\end{table}
\end{landscape}

The numerical result of the  form factor  in Eq.~(21) 
(as a example of numerical cross-check)
of Ref.~\cite{Hue:2017cph} is 
\begin{eqnarray}
\mathcal{F}_{21} &=& 
0.07245981549100559\times \frac{eg^2}{16 \pi^2}. 
\end{eqnarray}
We find a perfect agreement between 
the result in this paper with that  in Ref.~\cite{Hue:2017cph}.
\section{Conclusions}   
\noindent
The analytical results for the form factors presenting one-loop  
contributions of the $W$ boson  to the decay amplitude 
$H\rightarrow Z\gamma$ in the $R_\xi$ gauge have been collected.   
They are expressed as functions of  the  Passarino-Veltman functions 
that   numerical calculations are easily generated with {\tt LoopTools}.  In the limit of 
$d\rightarrow4$, we have shown that these analytic results 
are independent of the unphysical parameter $\xi$
and consistent with  those given in previous works. Numerical checks
for the $\xi$-independence of the form factors
has also discussed. The results are also in 
good stability with varying $\xi =0, 1, 100$ and 
$\xi \rightarrow \infty$.  We emphasize that  
the results in this paper will be applied to 
calculate one-loop contributions of new charged 
gauge bosons appearing in many  BSMs.  They can 
be used to cross-checks for consistence with  
well-known results given in the unitary gauge.   
This is  another indirect way to  confirm  the 
new goldstone boson couplings which  often 
have complicated forms in the BSMs. \\

\noindent
{\bf Acknowledgment:}~
This research is funded by Vietnam National Foundation
for Science and Technology Development (NAFOSTED) under 
the grant number $103.01$-$2019.346$.\\
\appendix 
\section{ Feynman rules and Feynman diagrams}         
In this Appendix, we list  Feynman rules needed for 
writing precisely all one-loop integrals contributing to the  decay amplitude
of  the process $H\rightarrow Z\gamma$. 
All propagators and related couplings are shown in 
Tables~\ref{Feynman rules table} and \ref{couplings table}, respectively. 
\begin{table}[h]
\centering
\scalebox{1.0}
{\begin{tabular}{ll}
\hline
 Types   & propagators\\
\hline
Goldstone boson
& $\dfrac{i}{p^2 - M_{\xi}^2}$ \\
Ghost 
& $\dfrac{i}{p^2 - M_{\xi}^2}$ \\
$W$ boson 
& $\dfrac{- i}{p^2 - M_W^2} 
\Bigg[ g^{\mu \nu} - (1 - \xi) \dfrac{p^\mu p^\nu}{p^2 - M_{\xi}^2}  \Bigg]$ \\
\hline
\end{tabular}}
\caption{Feynman rules involving the 
decay $H \rightarrow Z \gamma$ through 
$W$ boson loops in  the $R_\xi$ gauge.
\label{Feynman rules table}}
\end{table}
\begin{table}[h]
\begin{center}
\begin{tabular}{ll} 
\hline\hline
\textbf{Vertices} & \textbf{Couplings}   \\ \hline\hline
{\footnotesize 
$H \cdot W_\mu \cdot W_\nu$,
$H (p_1) \cdot W_\mu \cdot \chi (p_2)$,
$H \cdot \chi \cdot \chi$,
$H \cdot c \cdot c$}
& {\footnotesize 
$i g_{HWW} g_{\mu \nu}$,
$-i g_{HWX} (p_2 - p_1)_\mu$,
$-i g_{HXX}$,
$-i \xi g_{Hcc}$}  \\
{\footnotesize 
$A_\mu (p_1) \cdot W^+_\nu (p_2) \cdot W^-_\lambda (p_3)$,
$Z_\mu (p_1) \cdot W^+_\nu (p_2) \cdot W^-_\lambda (p_3)$}
& {\footnotesize 
$-i g_{AWW} \Gamma_{\mu \nu \lambda} (p_1, p_2, p_3)$,
$-i g_{ZWW} \Gamma_{\mu \nu \lambda} (p_1, p_2, p_3)$} \\
{\footnotesize 
$A_\mu \cdot A_\nu \cdot W^+_\alpha \cdot W^-_\beta$,
$Z_\mu \cdot A_\nu \cdot W^+_\alpha \cdot W^-_\beta$}
& {\footnotesize 
$-i g_{AAWW} S_{\mu \nu, \alpha \beta}$,
$-i g_{ZAWW} S_{\mu \nu, \alpha \beta}$}
\\
{\footnotesize 
$A_\mu \cdot W_\nu \cdot \chi$,
$Z_\mu \cdot W_\nu \cdot \chi$}
& {\footnotesize 
$-i g_{AWX} g_{\mu \nu}$,
$-i g_{ZWX} g_{\mu \nu}$} \\
{\footnotesize 
$A_\mu \cdot \chi (p_1) \cdot \chi (p_2)$,
$Z_\mu \cdot \chi (p_1) \cdot \chi (p_2)$}
& {\footnotesize 
$-i g_{AXX} (p_2 - p_1)_\mu$,
$-i g_{ZXX} (p_2 - p_1)_\mu$} \\
{\footnotesize 
$A_\mu \cdot A_\nu \cdot \chi \cdot \chi$,
$Z_\mu \cdot A_\nu \cdot \chi \cdot \chi$}
& {\footnotesize 
$i g_{AAXX} g_{\mu \nu}$,
$i g_{ZAXX} g_{\mu \nu}$} \\
{\footnotesize 
$H \cdot A_\mu \cdot W_\nu \cdot \chi$,
$H \cdot Z_\mu \cdot W_\nu \cdot \chi$}
& {\footnotesize 
$-i g_{HAWX} g_{\mu \nu}$,
$-i g_{HZWX} g_{\mu \nu}$} \\
{\footnotesize 
$A_\mu \cdot c \cdot c$,
$Z_\mu \cdot c \cdot c$}
& {\footnotesize 
$-i g_{Acc} p_\mu$,
$-i g_{Zcc} p_\mu$}
\\ \hline\hline
\end{tabular}
\end{center}
\caption{All couplings involving 
the decay $H \rightarrow Z \gamma$ through 
$W$ boson loops in the $R_\xi$ gauge. 
The notations 
defining  these couplings are:
$g_{HWW} = g M_W, 
 g_{HWX} = g / 2,
 g_{HXX} = g M_H^2 / (2 M_W), 
 g_{Hcc} = g M_W / 2, 
 g_{AWW} = e,
 g_{ZWW} = - g \cos\theta_W,
 g_{AAWW} = e^2,
 g_{ZAWW} = - e g \cos \theta_W,
 g_{AWX} = e M_W,
 g_{ZWX} = g M_Z \sin^2 \theta_W,
 g_{AXX} = e,
 g_{ZXX} = - g \cos 2\theta_W / (2 \cos \theta_W),
 g_{AAXX} = 2 e^2,
 g_{ZAXX} = - e g \cos 2\theta_W / (\cos \theta_W),
 g_{HAWX} = e g / 2,
 g_{HZWX} = g^2 \sin^2 \theta_W / (2 \cos \theta_W),
 g_{Acc} = e,
 g_{Zcc} = - g \cos \theta_W$.
The standard Lorentz tensors of the gauge boson self couplings are  $\Gamma_{\mu \nu \lambda} (p_1, p_2, p_3) = g_{\mu \nu} (p_1 - p_2)_\lambda + g_{\lambda \nu} (p_2 - p_3)_\mu + g_{\mu \lambda} (p_3 - p_1)_\nu$ and $S_{\mu \nu, \alpha \beta} = 2 g_{\mu \nu} g_{\alpha \beta} - g_{\mu \alpha} g_{\nu \beta} - g_{\mu \beta} g_{\nu \alpha}$.
\label{couplings table}}
\end{table}

\begin{figure}[h]
	\centering 
	\begin{tabular}{cc}
			\includegraphics[scale=0.5]{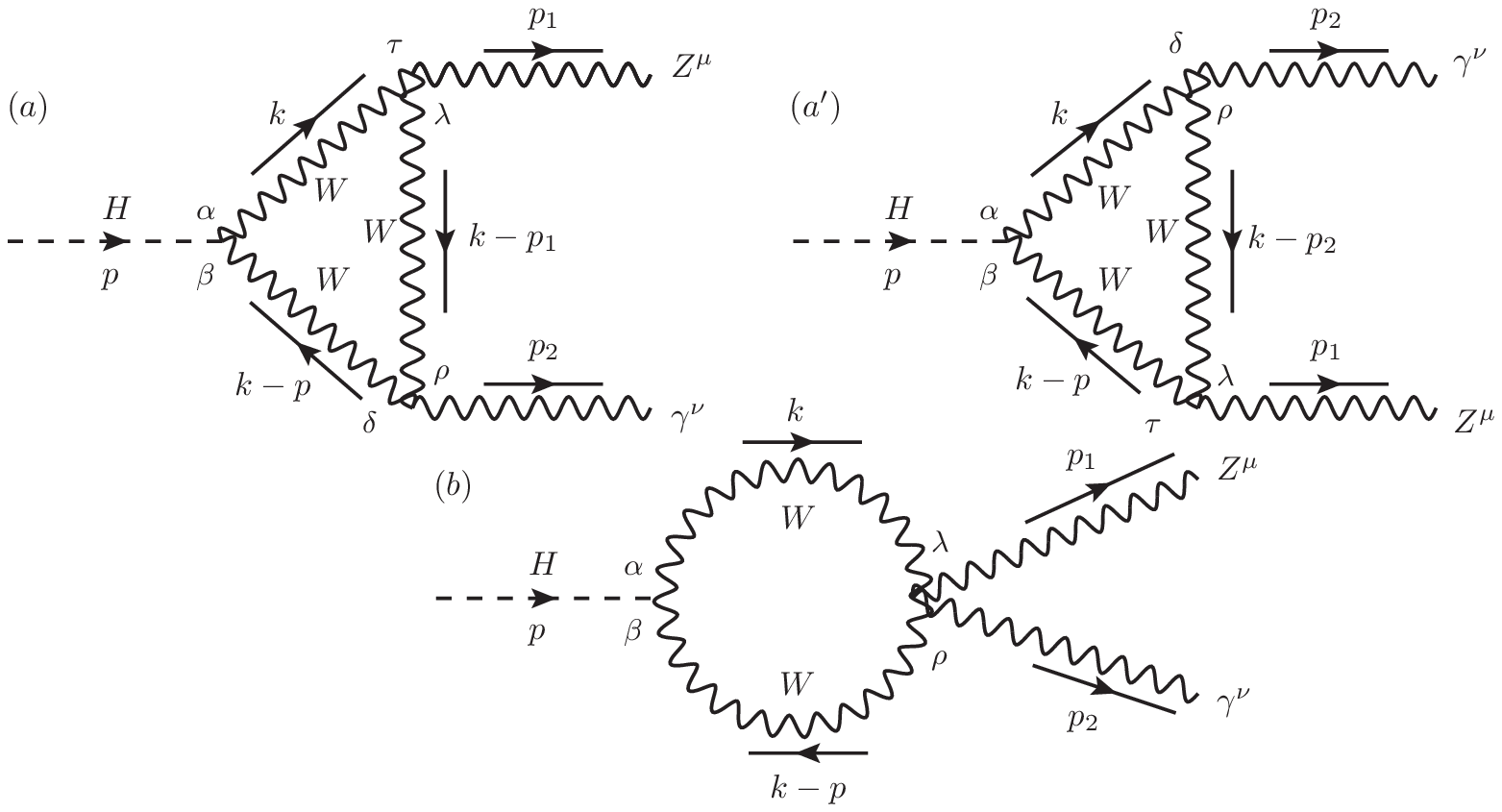}& 
				\includegraphics[scale=0.5]{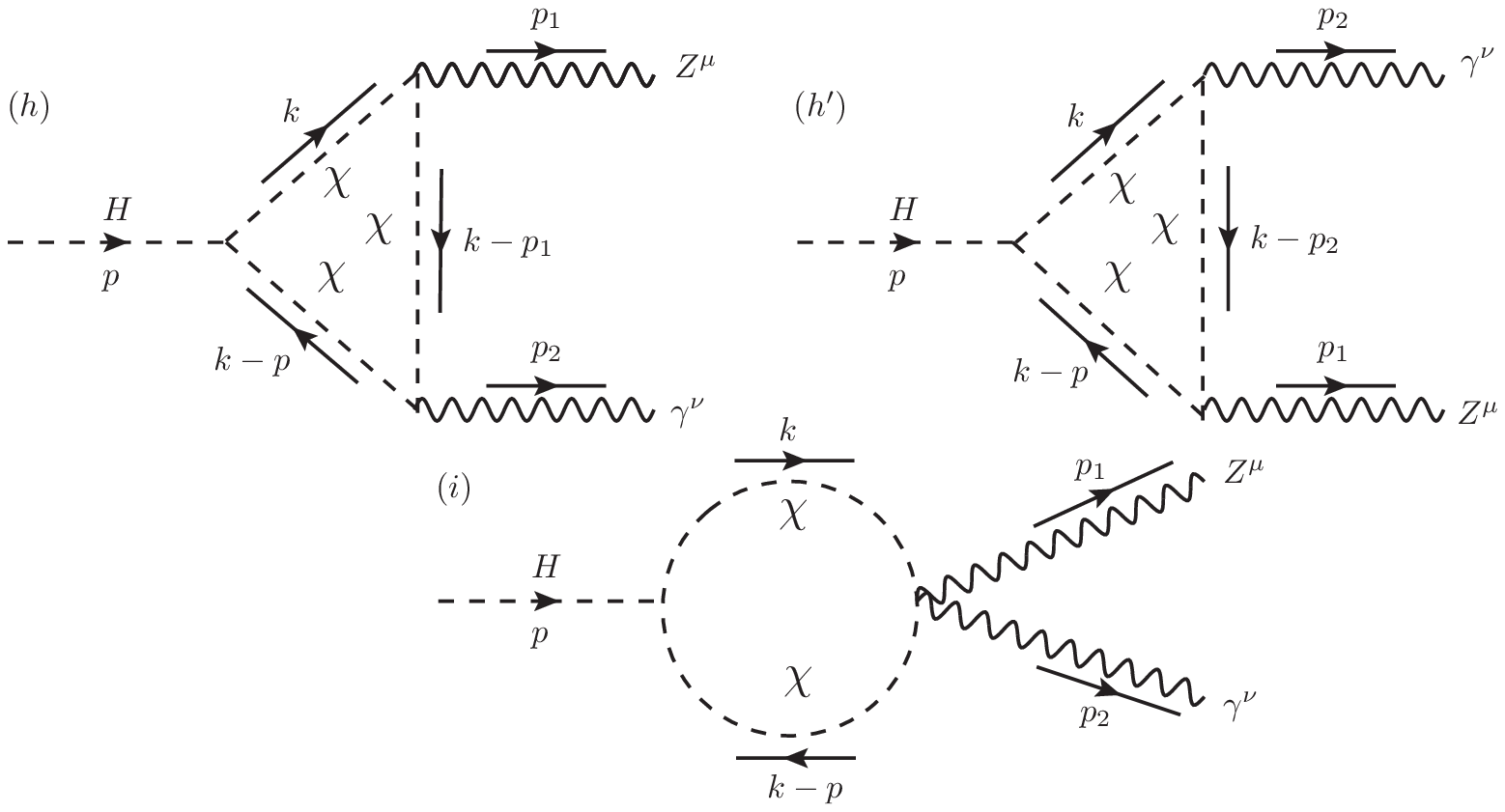}\\
					\includegraphics[scale=0.5]{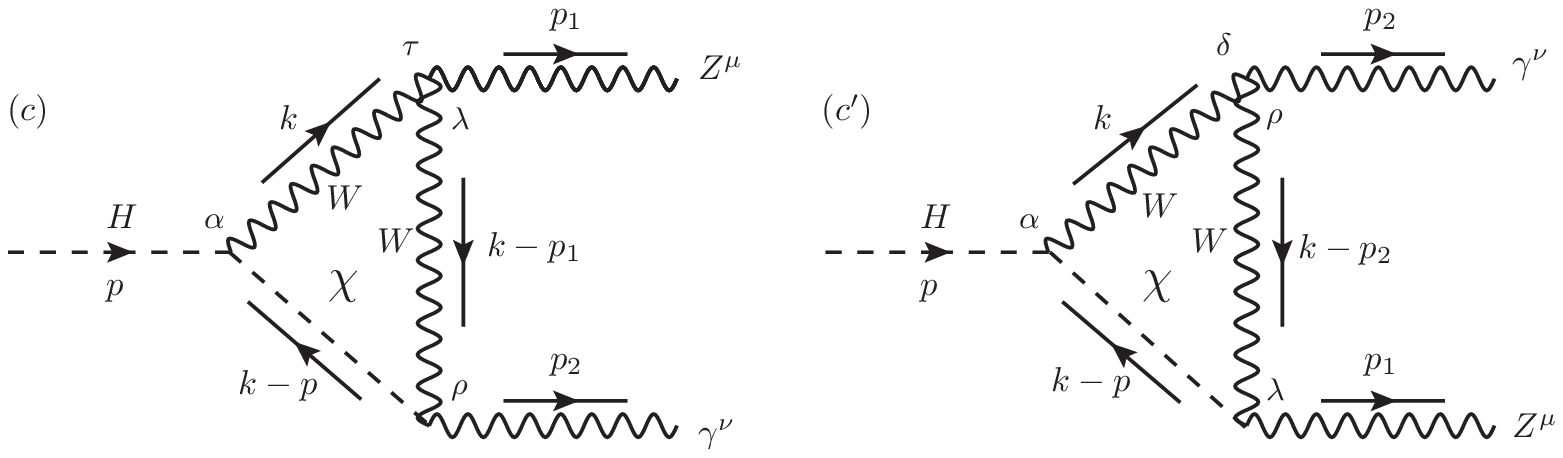}&
				\includegraphics[scale=0.5]{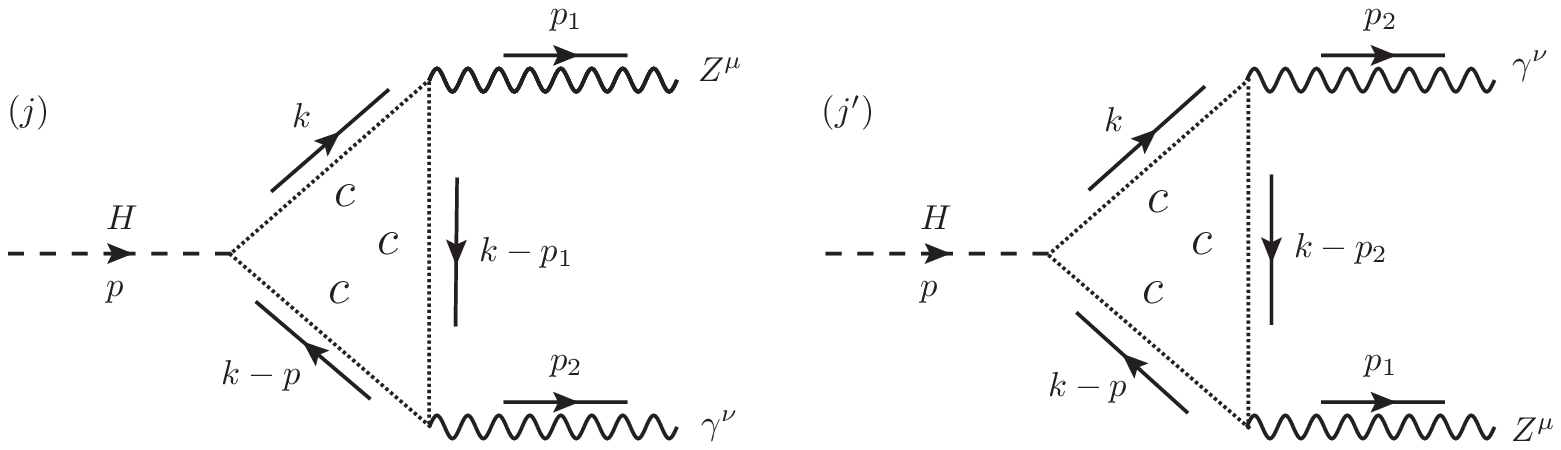}\\
				\includegraphics[scale=0.5]{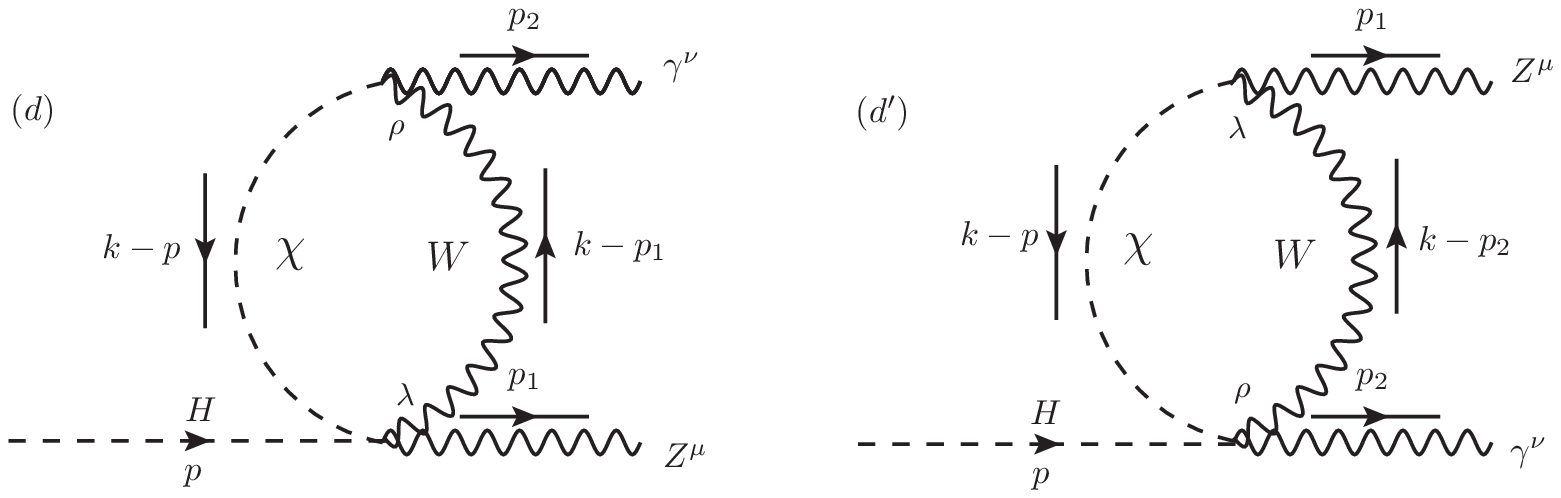}&
					\includegraphics[scale=0.5]{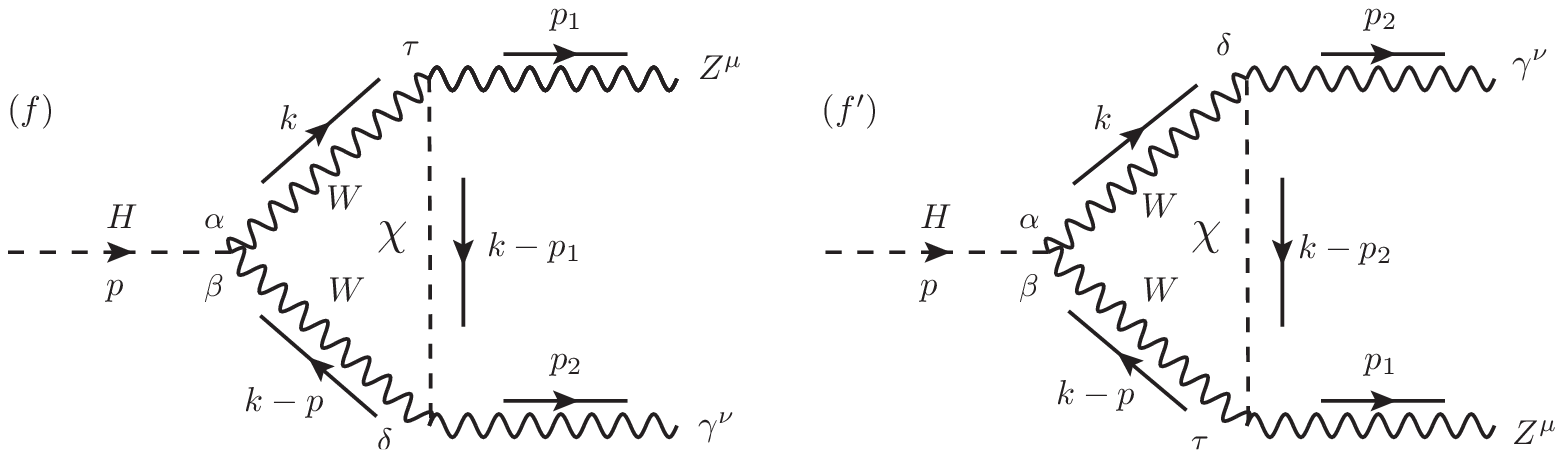}\\
			\includegraphics[scale=0.5 ]{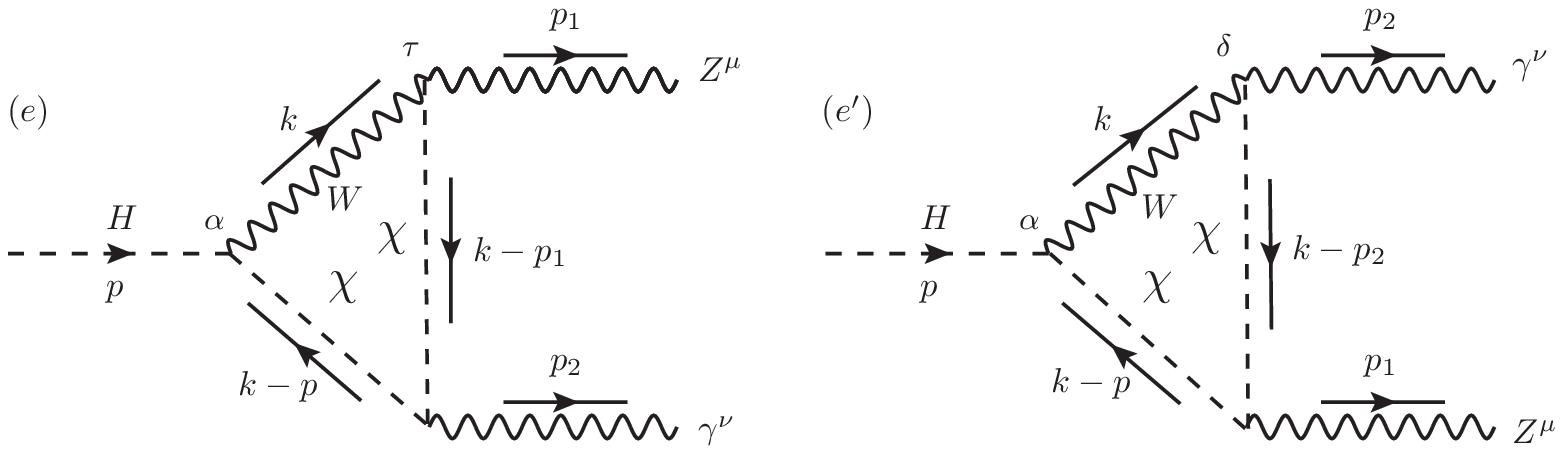}&
				\includegraphics[scale=0.5]{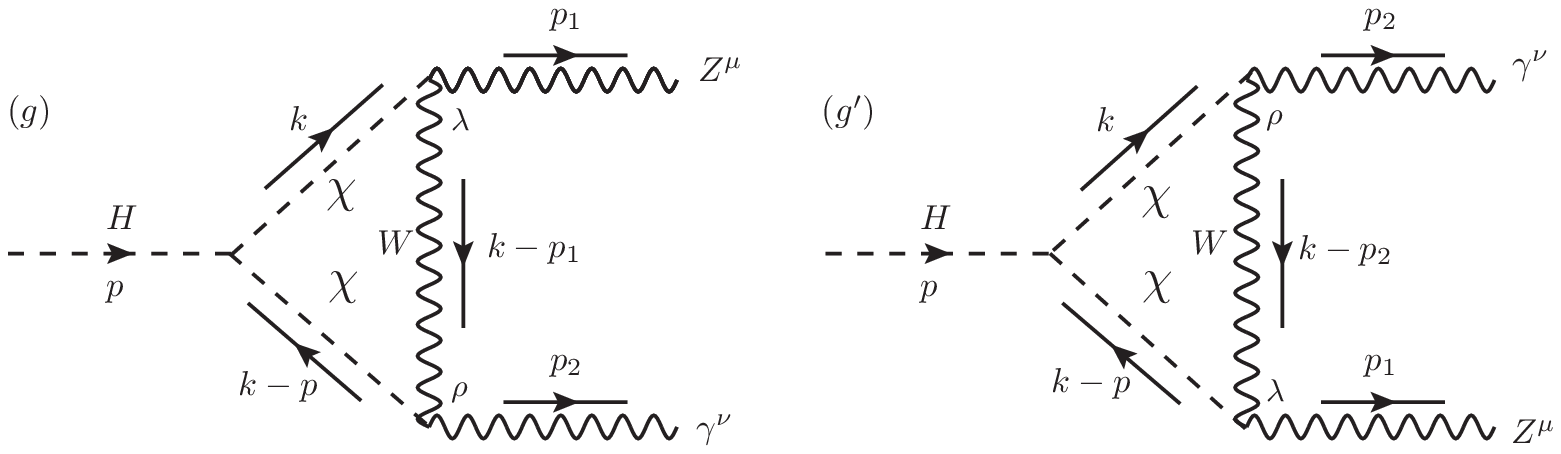}\\
			
			\end{tabular}
		\caption{\label{Wboson} Feynman diagrams of one-loop 
			$W$  boson contributions to $H\rightarrow Z\gamma$ in $R_{\xi}$.}
	\end{figure}

Feynman diagrams for one-loop contributions to the decay amplitude 
$H\rightarrow Z\gamma$ in $R_\xi$ gauge are 
plotted in Fig.~\ref{Wboson}. 

\end{document}